
\magnification=1200
\def\sect{\vskip 2mm \centerline}
\def\r{\hangindent=1pc  \noindent}
\def\cen{\centerline}
\def\v{\vskip 1mm}
\def\endpage{\vfil\break}
\def\kms{km s$^{-1}$}
\def\deg{$^\circ$}

\def\Msun{M_{\odot \hskip-5.2pt \bullet}}

\def\MH{m_{\rm H}}
\def\mH{$m_{\rm H}$}

\def\pcc{cm$^{-3}$}
\def\deg{$^\circ$}
\def\Deg{^\circ}

\def\htwo{H$_2$}
\def\pa{PASJ}

\def\pal{PASJ Letters}
\def\apj{ApJ}
\def\apjl{ApJ Letters}

\def\aa{A\&A}

\def\so{Sofue, Y.}

\def\fu{Fujimoto, M.}
\def\wa{Wakamatsu, K.}

\def\iaut{{\it Institute of Astronomy, University of Tokyo, Mitaka, Tokyo 181}}

\cen{\bf The Fate of the Magellanic Stream}
\cen{Yoshiaki SOFUE}
\cen{\iaut}

\v
\cen{(To appear in PASJ NOTE)}

\v\v\centerline{\bf Abstract} \v

We show that HI clouds in the Magellanic Clouds are stripped
by the ram-pressure due to the halo and disk gases of the Galaxy.
Molecular clouds are swept to the edge of the LMC, showing
an eccentric distribution.
The stripped HI clouds form a narrow band on the sky, and mimics
the Magellanic Stream, when the LMC takes a polar orbit.
We point out that the Magellanic Stream will fall into the Galaxy,
and will be finally accreted by the Galactic disk.
The accretion may cause warping of the inner gas disk of the Galaxy.
Some clouds are accreted by the nuclear disk, which could explain the
peculiar distribution and kinematics in the central region.
The stripping of the interstellar gas from the Magellanic Clouds will
rapidly change their morphological type into dwarf ellipticals.

\v

\sect{\bf 1. Introduction} \v

The dynamical evolution of the triple system of the Galaxy, the Large and Small
Magellanic Clouds has been extensively studied by test-particle simulations
(Fujimoto and Sofue 1976, 1977; Murai and Fujimoto 1980; Gardiner and Fujimoto
1993).
In these  models, the Magellanic Stream of HI gas (Mathewson et al 1974, 1977;
Mathewson and Ford  1984) has been treated as a tidal debris, which has been
torn off from the LMC and SMC through the gravitational interaction.
However, no hydro-dynamical effect has been taken into account.

The gas-dynamical interaction of the Magellanic Clouds with the intergalactic
and halo gases has been discussed only by a few authors:
A galactic-wake model was proposed by Mathewson et al (1977), and was examined
in a greater detail by Bregman (1979), who, however, found difficulty in
obtaining a satisfactory modeling of the Stream, and favored the gravitational
model.
On the other hand, the ram-pressure effect on gas clouds in the Magellanic
Clouds and Stream has not been thoroughly studied as yet.
In particular, the ram pressure would significantly affect the evolution of the
Magellanic Stream during its orbital motion around the Galaxy embedded in an
extended gaseous halo.

We have proposed a ram-pressure accretion model of intergalactic gas clouds by
galaxies with an extended gaseous halo, and showed that gaseous debris from
tidally disturbed galaxies are soon accreted by a nearby galaxy (Sofue and
Wakamatsu 1991, 1992, 1993).
We further performed a numerical simulation of the stripping-and-accretion
process of gas clouds from a companion onto its host galaxy, and applied the
result to the M31-M32-NGC 205 system (Sofue 1994).
In this paper, we attempt to examine the ram-pressure effect of the
circum-Galaxy gas on the dynamics of gas clouds in the Magellanic Cloud.
We also aim at investigating the fate of the Magellanic Stream in the future
during the interaction with the galactic halo and disk gases .

\sect{\bf 2. Basic Assumptions and Equations of Motion}\v

\sect{\it 2.1. HI and Molecular Clouds}\v

In this work, we treat ballistic orbits of HI and molecular clouds, which are
initially distributed in the LMC in the presence of the ram pressure due to the
intergalactic as well as the halo gases around the Galaxy.
Because of the larger effective cross section compared to the mass, HI clouds
are easily stripped from the companion than molecular clouds.
Stripped gas clouds can remain as intergalactic clouds when they are massive
enough, while less massive clouds are dissipated into the intergalactic space.
In the simulation, we assume that the HI clouds have  dimensions massive enough
to be gravitationally bound, balancing with its internal motion and/or rotation
(Table 1).
For molecular clouds we assume a size and mass of the same order as those of a
typical giant molecular cloud in our Galaxy, as shown in table 1, which are
also gravitationally balancing with the internal motion and/or rotation.

\cen{-Table 1-}

Although the cloud properties would vary during the interaction with the halo
and intergalactic gas, we here simply assume that their original properties are
kept unchanged during the simulation.
Although detailed analyses of such hydro-dynamical process within individual
clouds are beyond the scope of this paper, we make the following comment.
The internal motion such as turbulence within each cloud will be dissipated,
which would cause a collapse of the cloud.
However, instabilities on the cloud surface due to an interaction with the
intergalactic gas, such as the  Kelvin-Helmholtz instability, would excite
turbulence inside the cloud, and will act to maintain the internal motion.

\sect{\it 2.2. Equations of Motion and Potentials of the Galaxies}\v

We are mainly concerned with the accretion process of stripped clouds from the
LMC.
Since accretion orbits of individual clouds are not strongly dependent on their
initial internal motion within the LMC and SMC, we  consider the LMC only, and
its satellite (the SMC) is regarded to be a part of the envelope:
As the mass the LMC is much greater than that of the SMC, the interstellar gas
inside the SMC and the gaseous envelope surrounding the two Clouds are regarded
to be a single gaseous halo surrounding the LMC.

We adopt a simple ballistic model for ``test clouds", as it has been adopted in
the previous simulations (Sofue and Wakamatsu 1993; Sofue 1993).
The equation of motion for each cloud is written as;
$$
{{d^2{\bf r}}\over{dt^2}}
= \sum_{j=1}^2{{\partial \Phi_j}\over{\partial {\bf r}}}
- {{3 \rho({\bf r})}\over{4 R\rho_{\rm c}}} \Delta v {\bf \Delta v}, \eqno(1)
$$
with
$$
\Phi_1= \sum_{i=1}^3  {GM_i \over {\lbrace \varpi^2 + \left(a_i +
\sqrt{z^2+b_i^2}\right)^2 \rbrace^{1/2}}}, \eqno(2)
$$
and
$$
\Phi_2 = {GM_{\rm C} \over {\sqrt{r^2 + b_{\rm C}^2}}} \eqno(3)
$$
being the gravitational potentials for the Galaxy ($j=1$) and the companion
galaxy ($j=2$), which are approximated by the modified  Miyamoto and Nagai's
(1975) potential and the Plummer's law, respectively.
Here, $\varpi=(x^2+y^2)^{1/2}$,  and $M_i$, $a_i$ and $b_i$ are the mass, scale
radius and thickness for the $i$-th galaxy, respectively.
The vectors ${\bf v} = d{\bf r}/dt$ and  ${\bf r}=(x, y, z)$ represent the
cloud's velocity and position with respect to the center of the Galaxy, and
$\rho_{\rm c}$ and $\rho({\bf r})$ are the densities of the cloud and the
halo+disk gas of the Galaxy, respectively.
 ${\bf \Delta v} = {\bf v} - {\bf V}$ is the mutual velocity of the cloud and
diffuse gas.
For the Galaxy we assume three mass components: the central bulge ($i=1$); disk
($i=2$); and a massive halo ($i=3$).
The parameters are summarized in table 2.

\cen{-Table 2-}

We assume that the mass of the LMC is an order of magnitude smaller than the
main body of the Galaxy: $M_{\rm C}=0.1 M_2$, and that the center of mass of
the system coincides with the center of the Galaxy which is fixed to the origin
of the coordinates.
We take into account the dynamical friction on the companion's motion by the
massive halo.
The equation of motion of the center of the LMC is written as
$$
{{d^2{\bf r}}\over{dt^2}}
= {{\partial \Phi_1}\over{\partial {\bf r}}}
- k \rho_{\rm MH}(r) M_{\rm C}{{\bf v}\over v},
\eqno(4)
$$
where $M_{\rm C}$ is the mass of the companion.
The second term represents the dynamical friction due to the massive halo,
which is assumed to be at rest.
The density of the massive halo is assumed to be inversely proportional to the
square of $r$: $\rho_{\rm MH}(r)=\rho_{\rm MH0}(r/100~{\rm kpc})^{-2}$, with
$\rho_{\rm MH0}$ being a constant.
The variable  $k$ represents the coefficient of the dynamical friction.
Although coefficient $k$ is actually a slowly varying function of velocity and
mass (Tremain 1976),  we here assume it to be constant.
We took a value for $k \rho_{\rm MH0}$ so that the acceleration by the second
term becomes equal to 0.005 times the gravitational acceleration by the first
term  when the companion galaxy is at a distance of 100 kpc from the center.

\sect{\it 2.3. Models for Gaseous Disk, Halo and Intergalactic Gas}\v

A hot tenuous halo of extended gas around the Galaxy has been theoretically
predicted (Spitzer 1956) and was discovered by IUE observations (Savage and de
Boer 1979).
In this paper, we consider three components of extended gas around the Galaxy:
the disk gas, halo, and the intergalactic diffuse gas.
We assume  a simple density distribution around the Galaxy as represented by
the following equation:
$$
\rho({\rm r})
=\rho(\varpi, z)
= \rho_0
+{\rho_{\rm H} \over{(\varpi/\varpi_{\rm H})^2 + (z/z_{\rm H})^2 + 1 }}
+{\rho_{\rm D} \over{(\varpi/\varpi_{\rm D})^2 + (z/z_{\rm D})^2 + 1 }}
, \eqno(5)
$$
where $\rho_{\rm H}$, $\varpi_{\rm H}$ and $z_{\rm H}$ are parameters
representing the distribution of the halo gas density,
$\rho_{\rm D}$, $\varpi_{\rm D}$ and $z_{\rm D}$ are those for the disk
component,
and $\rho_0$ is the  intergalactic gas density (Sofue and Wakamatsu 1993).
Values of the parameters are given in Table 2.

Although little is known about the rotation of the halo gas, we here assume
that the halo gas is rotating around the $z$ axis with its centrifugal force
balancing  the galaxy's gravity toward the $z$-axis, following the assumption
taken by Sofue and Wakamatsu (1993).
We assume that the gas is in a hydrostatic equilibrium in the $z$-direction, so
that $V_z=0$, and pressure gradient in the $\varpi$ direction is neglected.
This assumption would be too simplified and the neglect of pressure gradient in
the $\varpi$ direction would result in an overestimation of the rotation speed
of the halo gas.

\sect{\it 2.4. Orbits of the LMC and the Initial Conditions}\v

The orbit of the Magellanic System has been extensively investigated by
numerical integrations of the orbits toward the past, so that the present
binary state has been guaranteed for the past ten Gyr as well as that the
Magellanic Stream can be reproduced by the tidal debris (Fujimoto and Sofue
1976, 1977; Murai and Fujimoto 1980).
These computations indicated that the LMC is rotating around the Galaxy along a
polar orbit that is counter-clockwise as seen from the Sun.
In this paper, we adopt the polar and semi-polar (high-inclination) orbits, and
examine both clockwise and counter-clockwise orbits.

We solve the differential equations by using the  Runge-Kutta-Gill method.
The time step of integration was taken to be smaller than 0.005 times the
dynamical time scale of each test cloud.
The initial position of the LMC is taken at a sufficiently large distance on
the $z$ or $x$ axis, and the initial velocity is given so that it takes a
(semi-)polar orbit.
The initial velocity is given so that the distance and radial velocity at the
present position approximately coincides with the observed distance (52 kpc)
galactocentric radial velocity (51 \kms) of the LMC (Allen 1973).
Initial values of the coordinates and velocity of the LMC are given in figure
captions for individual results (Fig. 2 - 5).

$N(=50)$ interstellar clouds are initially distributed at random in the
companion within radius $R (=10$ kpc) with a velocity dispersion $\sigma_v (=
50$ \kms), so that the ensemble of test clouds are maintained to be a spherical
system.
This might be replaced with a rotating disk of a similar size.
Since the initial distribution within the companion little affects the
accretion process after stripping (Sofue 1993),  we adopt here simply a
spherical distribution.

\sect{\bf 3. Results}\v

Fig. 1a shows the result of the numerical integration for a counter-clockwise
polar orbit, which mimics the orbit taken by Fujimoto and Sofue (1976, 1977).
Fig. 1b is the same but in a stereo-gram, and Fig. 1c is an enlargement of Fig.
1a.
The upper panel shows the $(x,z)$ plane, where the Galaxy lies edge-on, and we
can obtain a rough idea of the distribution of clouds when they are projected
on the sky in the $(l, b)$ coordinates.
The lower panel is a projected view on the $(x, y)$ plane (the Galactic plane).
Fig. 2, 3 and 4 show the cases for clockwise orbits, where Fig. 2 is for a
polar orbit, Fig. 3 for a semi-polar prograde (direct) orbit, and Fig. 4 for a
semi-polar retrograde orbit with respect to the galactic rotation.
Fig. 5 plots the results on the sky as seen from the Sun.

The simulation shows that the orbits of the individual gas clouds change
drastically when they cross the galactic plane, where the clouds suffer from
the strongest ram-braking due to the rotating disk.
In all cases, the HI test clouds are entirely stripped, and the molecular
clouds are partly stripped.
The ``efficiency'' of stripping of HI clouds may be too high compared to the
observations: this is because the individual clouds suffers from the ram
pressure, even when they are distributed in the LMC.
In reality, however, clouds within the LMC are ``shielded'' from the halo and
intergalactic gas by many other clouds and by the diffuse component, so that
the effective ram would be smaller.
Therefore, the real stripping of clouds will be more gentle than the simulation
shows.
The decrease in the apogalactic distance due to the dynamical friction is
gradual, and is negligible during the period of the present simulations.

\cen{ -- Fig. 1 to 5 --}

\sect{\it 3.1. Eccentric Distribution of Clouds in the LMC}

During the passage of the galactic plane, HI clouds in the LMC suffer from the
strongest ram pressure.
The clouds attain an eccentric distribution, being accumulated in the
down-stream side of the LMC, and are swept away from the central region.
After the passage, the HI clouds are almost entirely stripped.
On the other hand, stripping of molecular clouds occurs more gently, and many
clouds survive the stripping:
most of the molecular clouds are only accumulated in the edge of the LMC
without being stripped, except for outer clouds which are stripped.
They attain an eccentric distribution, and begin to oscillate in the potential
of the LMC.
Such an eccentric distribution is indeed observed both for CO (Cohen et al.
1988) and HI (Mathewson and Ford 1984).
We stress that the eccentric distribution of gas displaced from the stellar
body cannot be reproduced by the gravitational model.

\sect{\it 3.2. The Magellanic Stream}

The stripped HI clouds  form a tail trailing from the LMC.
As they leave the LMC and are attracted by the Galaxy's potential, they get
more elongated.
Since they begin to rotate around the Galaxy at a higher angular velocity than
the LMC, the clouds form a ``leading stream'' on the sky.
The calculated stream for the counter-clockwise orbit reproduces qualitatively
the observed Magellanic Stream (Mathewson and Ford  1984), when the LMC reaches
the present position (Fig. 1).
It is conspicuous that the stream becomes as narrow as about 4 kpc across,
getting narrower and longer as they approach the Galaxy.
This stream is, however, finally accreted by the galactic disk.

In order to compare the simulation with the observed HI stream (Mathewson and
Ford 1984), we plotted the calculated results in the galactic coordinates in
Fig. 5, where the ram pressure is assumed to be a half that assumed for Fig. 1.
The simulation for Fig. 5a reproduces qualitatively the observed Stream.
However, the detail is not well reproduced, which will be due to the too
simplified assumption of size and mass of clouds, as well as due to ignoring
the shielding effect of clouds inside the LMC by other clouds and diffuse
component.
If we take into account  a wider range of effective-ram force on clouds, the
stream would be more elongated and will better reproduce the observations.
We point out that not only the tail, but also ``pre-stream'' clouds, which are
observed at ($l, b) \sim (300\Deg, -10\Deg)$, are simulated.
It is also interesting to note that the simulated stream is not perfectly
aligned along the LMC's projected orbit, but is displaced toward the direction
of the SMC by about 10\deg, which is indeed observed in the HI gas
distribution.

\sect{\it 3.3. The Fate of the Magellanic Stream: Accretion onto the Galaxy}

In all cases of the simulations, stripped HI clouds are finally trapped by the
Galaxy, and infall toward the galactic disk along polar ``spiral'' orbits.
The accretion occurs within $\sim 10^9$ yr after the clouds have left the LMC's
gravity.
For the polar and prograde (direct) orbits, regardless clockwise or
anti-clockwise,  HI clouds are finally accreted by the galactic disk, and form
a ring of a radius about 10 kpc.
For the retrograde orbits, on the other hand, the infalling clouds hit the
central region of the Galaxy, and form a compact disk (ring) of a few kpc
radius around the nucleus.

The accretion of molecular clouds is much slower, and they remain as
intergalactic or intra-halo gaseous debris for a longer time than HI.
Finally, however, they are also accreted by the Galaxy in several $10^9$ yr.
During the accretion, some molecular clouds take peculiar polar orbits, and hit
the central region of the Galaxy.

\sect{\bf 4. Discussion}\v

We have shown that gas clouds in the LMC are stripped by ram-pressure of the
halo gas, and that the Magellanic Stream will be accreted by the Galaxy in a
few Gyr.
We discuss some implications of the results for the gas dynamics in the Galaxy
and for the evolution of the Galaxy-Magellanic Clouds System.

\sect{\it 4.1. Infalling Streams and High-velocity Clouds}\v

We may suppose that the stripping-and-accretion of HI and molecular clouds
occurred recurrently during the past tidal as well as gas-dynamical interaction
of the LMC and the Galaxy, particularly when the LMC crossed the galactic
plane.
We may, therefore, expect that a larger number of Magellanic-Stream-like clouds
have fallen into the Galaxy in the past, which  would be observed as
high-velocity HI clouds (e.g., van Woerden et al. 1985).
As the simulation indicates, directions and velocities of the infalling clouds
near the galactic plane are not apparently related to the orbit of the LMC
because of the larger friction of the disk gas of the Galaxy.

\sect{\it 4.2. Warp and Peculiar Kinematics of the Inner Galactic Disk}

The falling gas clouds hit the central region of the Galaxy, when the LMC's
orbit is retrograde.
If we take into account the finite amount of the angular momentum of the
original gas disk in the Galaxy, particularly of the inner disk, the infall
will cause a significant change of the angular momentum of the gas disk (Sofue
and Wakamatsu 1993).
It is quite possible that such infalling streams have hit the central region of
the Galaxy in the past, which caused a significant change of the rotation
characteristics of the inner disk.
This could explain the tilt of the inner 1-kpc HI disk, which is warping by
about 10\deg from the galactic plane (Burton and Liszt 1978).
The angular momentum of of the inner disk of 1 kpc radius is of the order of $
\sim M r V \sim 1.2 \times 10^{65}$ g cm$^2$ s$^{-1}$ for $M\sim 10^8\Msun,~
r\sim  1$ kpc and $V\sim 200$ \kms.
If the infalling stream had a mass of $\sim 10^7\Msun$ and has hit the inner
disk at a velocity of 200 \kms perpendicular to the disk, the additional
angular momentum, $AM \sim 10^{64}$ g cm$^2$ s$^{-1}$, would be sufficient to
change the angular momentum vector of the inner disk by about ten degrees.

The infall of clouds into the more central region would further cause peculiar
kinematics and distribution of the molecular gas in the nuclear disk.
Indeed, the molecular gas in the central few hundred pc of our Galaxy exhibits
a highly eccentric (asymmetric) distribution, and shows even ``forbidden''
velocities (e.g. Bally et al. 1987).
We point out that the infalling high-velocity clouds would produce ``sprays''
of gas, which vertically expand from the nuclear disk toward the halo, as well
as shock waves inside the disk.
These appear to be indeed observed in the central few-hundred pc region of the
Galaxy in the form of various exotic features seen in radio and molecular lines
(e.g., Sofue 1989).

\sect{\it 4.3. Evolution of the Magellanic Clouds into Early-type Galaxies}\v

The stripping of gas clouds from a companion and their accretion onto the
Galaxy applies generally to any multiple-galaxy systems, and should play a
significant role in the evolution of the smaller-mass galaxies.
In particular, the selective and one-way transfer of gas will result in a rapid
evolution of the companion:
The companion gets more gas deficient and, therefore, redder (early type),
whereas its host galaxy gets gas-richer and bluer (later type).
We stress that the gas transfer between interacting galaxies occurs much more
rapid than the stellar mass transfer due to the gravitational effect such as
the dynamical friction.
We, therefore, predict that the LMC and SMC will evolve from the present young
and blue irregular type into gas-poor dwarf ellipticals in the future, as far
as the two Clouds are bound to the Galaxy.
Moreover, the binary state of the LMC and SMC will be destroyed by the stronger
tidal force, as they approach the Galaxy due to the dynamical friction.
Hence, the future state of the Magellanic Clouds will mimic that of the dwarf
elliptical companions (M32 and NGC 205), which are already gas deficient,
around M31 (Sofue 1994).

\v\v
\sect{\bf References}\v

\r Allen, C. W. 1973, {\it Astrophysical Quantities} (University of London, The
Athlone Press, London), Ch. 14.

\r Bally, J., Stark, A. A., Wilson, R. W., Henkel, C. 1987, {\it Ap. J. Suppl.
}, {\bf 65}, 13.

\r Bregman, J. N. 1989, \apj, 229, 514.

\r Burton, W. B., and Liszt, H. S. 1978, \apj, l225, 815.

\r Cohen, R. S., Dame, T. M., Garay, G., Montani, J., Rubio, M., and Thaddeus,
P. 1988, \apjl, 331, L95.

\r \fu, and \so 1976, \aa, {\bf 47}, 263.

\r \fu, and \so 1977, \aa, {\bf 61}, 199.


\r Mathewson, D. S., and Ford, V. L. 1984, in {\it Structure and Evolution of
the Magellanic Clouds, IAU Symp. No. 108}, ed. S. van den Bergh and K. S. de
Boer (Reidel Publishing Co., Dordrecht), p. 125.

\r Mathewson, D. S., Cleary, M. N., and Murray, J. D. 1974, \apj, 190, 291.

\r Mathewson, D. S., Schwarz, M. P., and Murray, J. D. 1977, \apjl, 217, L5.

\r Miyamoto, M., and Nagai, R. 1975, \pa, {\bf 27}, 533.

\r Murai, T., and \fu\ 1980, \pa, 32, 581.

\r Savage, B. D., and de Boer, K. S. 1979, \apj, .

\r Spitzer, L. 1956, \apj, 124, 20.

\r \so\ 1989, in {\it The Galactic Center (IAU Symp. No. 136)}, ed. M.Morris,
(Reidel Publ. Co., Dordrecht), p.213

\r \so\ 1994, \apj, March issue in press.

\r \so, and \wa\ 1991, \pal, 43, L57.

\r \so, and \wa\ 1992, \pal,  44, L23.

\r \so, and \wa\ 1993, \aa, 273, 79.

\r Tremain, S. D. 1976, \apj, 203, 72.

\r van Woerden, H., Schwarz, U. J., and Hulsbosch, A. N. M. 1985, in {\it The
Milky Way Galaxy}, ed. H. van Woerden and R. J. Allen (D. Reidel Publ. Co.,
Dordrecht), p. 387.

\endpage

\settabs 2 \columns \v
\cen{Table 1: Parameters for gaseous components.}
\v
\hrule \vskip 0.5mm \hrule
\v
\+ Intergalactic gas \cr
\+ ~~~~~~~$\rho_0$ 	~\dotfill~ & $10^{-5}\MH$ \pcc \cr
\v
\+ Halo of the Galaxy \cr
\+ ~~~~~~~$\rho_{\rm H}$ 	~\dotfill~ &  0.01 \mH \pcc\cr
\+ ~~~~~~~$\varpi_{\rm H}$ 	~\dotfill~ & 15 kpc \cr
\+ ~~~~~~~$z_{\rm H}$ 	~\dotfill~ & 10 kpc \cr
\v
\+ Disk of the Galaxy \cr
\+ ~~~~~~~$\rho_{\rm D}$ 	~\dotfill~ & 1 \mH \pcc \cr
\+ ~~~~~~~$\varpi_{\rm D}$ 	~\dotfill~ & 10 kpc \cr
\+ ~~~~~~~$z_{\rm D}$ 	~\dotfill~ & 0.2 kpc \cr
\v
\+ Molecular cloud\cr
\+ ~~~~~~~$\rho_{\rm cloud}$ 	~\dotfill~ & $100$ \htwo \pcc \cr
\+ ~~~~~~~$R$	~\dotfill~ & 30 pc \cr
\+ ~~~~~~~$m=(4\pi/3)R^3\rho_{\rm HI}$ ~\dotfill~ & $ 1.32\times10^5 \Msun$
\cr\+ ~~~~~~~$\sigma_{\rm cloud}$	~\dotfill~ & 4.36 \kms \cr
\v
\+ HI cloud\cr
\+ ~~~~~~~$\rho_{\rm HI}$ 	~\dotfill~ & $1$ \mH \pcc \cr
\+ ~~~~~~~$R$	~\dotfill~ & 500 pc \cr
\+ ~~~~~~~$m=(4\pi/3)R^3\rho_{\rm HI}$ ~\dotfill~ & $ 3.05\times10^6\Msun$  \cr
\+ ~~~~~~~$\sigma_{\rm HI}$	~\dotfill~ & 5.11 \kms \cr
\v
\hrule

\endpage
\settabs 6 \columns \v
\cen{Table 2: Parameters for the gravitational potentials of the Galaxy and
LMC.}
\v
\hrule \vskip 0.5mm \hrule
\v
\+  ~~~ $ i $ &Mass component & & $M_i(\Msun)$ & $a_i$ (kpc) & $b_i$ (kpc) \cr
\v
\hrule
\v
\+ The Galaxy$^\dagger$ \cr
\+ ~~~1   \dotfill  &Central bulge &  & $2.05\times 10^{10}$ & 0 & 0.495 \cr
\+ ~~~2 \dotfill  & Disk &  & $2.547 \times 10^{11}$ & 7.258 &  0.520 \cr
\+ ~~~3  \dotfill  & Massive halo & & $3 \times 10^{11}$ & 20 & 20 \cr
\v
\+ Companions$^\ddagger$ \cr
\+ ~~~LMC \dotfill  & Spheroid  &\hskip 20mm $M_{\rm C}=0.1\times M_2$ && 0 & 2
\cr
\v
\hrule
\v
\noindent $\dagger$ Miyamoto-Nagai's (1975) potential with a modified massive
halo.

\noindent $\ddagger$ Plummer's potential. For a convenience to trace the global
behavior of stripped gas clouds, we consider the LMC+SMC system as a single
object represented by LMC.

\endpage

\noindent{Figure Captions} \vskip 3mm

\r Fig. 1a: Ram-pressure stripping of HI (large dots) and molecular (small
dots) clouds from the LMC, and their accretion onto the Galaxy for a
counter-clockwise (as seen from the Sun) polar orbit.
The initial condition is $(x, y, z)=(-100, 0, 100)$ kpc and $(v_x, v_y,
v_z)=(0, 0, -80)$ \kms.
The upper panel is the projection onto the $(x, z)$ plane (the Galaxy is edge
on), and the lower panel onto the $(x,y)$ plane (Galactic plane).
The Galaxy is rotating clockwise on this $(x,y)$ plane.
The square represents the position of the LMC plotted every 0.1 Gyr.
Stripped HI clouds reproduce the Magellanic Stream, while molecular clouds show
an eccentric distribution within the LMC.

\r Fig. 1b: Same as a, but in a stereogram.

\r Fig. 1c: Same, but the scale is doubled.

\v\r Fig. 2: Same as Fig. 1, but for a clockwise polar orbit with the initial
condition of $(x, y, z)=(-100, 0, 0)$ kpc and $(v_x, v_y, v_z)=(0, 0, 100)$
\kms, and the peri-G is 40 kpc.
The HI tail also resembles the Magellanic Stream.

\v\r Fig. 3: Same as Fig. 3, but for a prograde orbit with respect to the
galactic rotation.
The initial condition is $(x, y, z)=(-100, 0, 0)$ kpc and $(v_x, v_y, v_z)=(0,
50, 80)$ \kms.

\v\r Fig. 4a: Same as Fig. 3, but for a retrograde orbit.
The initial condition is $(x, y, z)=(-100, 0, 0)$ kpc and $(v_x, v_y, v_z)=(0,
-50, 100)$ \kms.
The accreted HI clouds hit the central region of the Galaxy.

\r Fig. 4b: Same as Fig. 4a, but in a stereogram.

\r Fig. 5: Plots of the results in Fig. 1 to 4 on the $(l, b)$ coordinates,
respectively in panels (a) to (d).
Note that the ram force for HI clouds in (a) is assumed to be a half that
assumed for Fig. 1.
The northern and southern hemispheres are projected on the galactic plane.
The galactic poles are at the center, and the outer circle represents the
galactic equator.
The small circles represent the LMC's position, and the diameter is inversely
proportional to the distance from the Sun.

\bye